\documentclass[fleqn,11pt]{article}

\usepackage[cp1251]{inputenc}
\usepackage[english]{babel}
\usepackage{amsfonts,amsmath,amssymb}
\usepackage{latexsym}

\usepackage[colorlinks=true]{hyperref}

\usepackage{amsfonts,amssymb,cite}
\usepackage{graphicx}

\def\noi{\noindent}
\def\jnumber#1#2{\thispagestyle{empty} \noi\unitlength=1mm
    	\begin{picture}(178,10)
            \put(177,15){\llap{\large\it Grav. Cosmol. No.\,#1, #2}}
                    \end{picture}}

\newcommand{\Title}[1]{\noi {{\Large\bf #1}}\\[1ex]}

\def\Aunames#1{\noi{\bf #1}}
\def\au#1{${}^{#1}$}
\def\Addresses#1{\medskip\noi \protect
	\begin{description}\itemsep -3pt {\it #1} \end{description}}
\def\adr#1#2{\item[${}^{#1}$]{\it #2}}

\newcommand{\Abstract}[1]{\vskip 2mm \begin{center}
        \parbox{16.4cm}{\small\noi #1} \end{center}\medskip}

\def\email#1#2{\footnotetext[#1]{e-mail: #2}\addtocounter{footnote}{1}}

\def\nqq{\hspace*{-2em}}

\def\cm{\hspace*{1cm}}

\usepackage{color}

\def\Funding#1{\subsection*{Funding} #1}

\def\Jl#1#2{#1 {\bf #2},\ }

\def\ApJ#1 {\Jl{Astroph. J.}{#1}}
\def\CQG#1 {\Jl{Class. Quantum Grav.}{#1}}
\def\DAN#1 {\Jl{Dokl. AN SSSR}{#1}}
\def\GC#1 {\Jl{Grav. Cosmol.}{#1}}
\def\GRG#1 {\Jl{Gen. Rel. Grav.}{#1}}
\def\IJMPD#1 {\Jl{Int. J. Mod. Phys. D}{#1}}
\def\JETF#1 {\Jl{Zh. Eksp. Teor. Fiz.}{#1}}
\def\JETP#1 {\Jl{Sov. Phys. JETP}{#1}}
\def\JHEP#1 {\Jl{JHEP}{#1}}
\def\JMP#1 {\Jl{J. Math. Phys.}{#1}}
\def\NPB#1 {\Jl{Nucl. Phys. B}{#1}}
\def\NP#1 {\Jl{Nucl. Phys.}{#1}}
\def\PLA#1 {\Jl{Phys. Lett. A}{#1}}
\def\PLB#1 {\Jl{Phys. Lett. B}{#1}}
\def\PRD#1 {\Jl{Phys. Rev. D}{#1}}
\def\PRL#1 {\Jl{Phys. Rev. Lett.}{#1}}

\def\lal{&&\nqq {}}

\def\beq{\begin{equation}}
\def\eeq{\end{equation}}
\def\bear{\begin{eqnarray}}
\def\bearr{\begin{eqnarray} \lal}
\def\ear{\end{eqnarray}}
\def\earn{\nonumber \end{eqnarray}}

\def\nnn{\nonumber\\ \lal }

\def\yy{\\[5pt] {}}

\begin{document}
\twocolumn[
\jnumber{issue}{year}

\Title{Evaluation of the magnetic field of hot Jupiters \yy 
	   within the geometric approach}
	   		
\Aunames{Babenko I.A.,\au{a,1} Zhilkin A.G.\au{b,2}}

\Addresses{
\adr a {Institute of Gravitation and Cosmology, RUDN University, 117198 Moscow, Russia}
\adr b {Institute of Astronomy, Russian Academy of Sciences, 119017 Moscow, Russia}
}


\Abstract
	{Theoretical and experimental foundations of the hypothesis about the origin of the magnetic fields of the Earth and other astrophysical objects, proposed in the early 20th century by W. Sutherland, A. Einstein, and independently by Yu.S. Vladimirov, are discussed in the paper. According to this hypothesis, the electric charges of the electron and proton slightly differ in magnitude, leading to the emergence of a magnetic field in rotating astrophysical objects. The theoretical justification of the Sutherland–Einstein hypothesis is presented in a simplified version of the 6D Kaluza--Klein theory, taking into account the consequences of the Kerr--Newman metric. The analysis shows that a fundamental dipole-type magnetic field should arise around any massive rotating body. However, in real astrophysical objects, such a field is largely screened and distorted by induced charges and currents. As an application, we consider the problem of determining the magnetic fields of hot Jupiters, since the strong tidal effects in these giant exoplanets should result in approximately similar screening mechanisms.}
\medskip
]%

\email 1 {inna.babenko@list.ru\\ \cm (Corresponding author)}
\email 2 {zhilkin@inasan.ru}

\section{Introduction}

One of the fundamental questions in modern astrophysics is understanding the physical nature of the magnetic fields of astrophysical objects. Throughout the 20th century, numerous theoretical efforts were made to explain the origin and properties of the magnetic fields of the Earth, the Sun, and, based on the resulting model, to provide explanations for the magnetic fields of other astrophysical objects. Attempts to explain the nature of terrestrial magnetism have been made for a long time, beginning with the works of W. Gilbert (1600), where the Earth was interpreted as a large permanent magnet.

To date, the most developed approach is the dynamo theory. In various astrophysical objects, either laminar (due to non-axisymmetric motions) \cite{Braginsky1964} or turbulent \cite{Parker1979, Moffat1978, Vainshtein1980} dynamo mechanisms can operate. Field amplification occurs under the condition that the average helicity of non-axisymmetric and turbulent plasma motions lacks mirror symmetry relative to the equatorial plane of the rotating body. This means that the number of right-handed vortices is not equal to the number of left-handed ones, as the Coriolis force provides additional twisting of individual vortices.

The main difficulties of dynamo theory are related to solving the complete system of magnetohydrodynamic equations (see e.g. \cite{Ruzmaikin1988}). To approximately solve this challenging problem, researchers resort to a number of simplifications, leading to a large variety of models \cite{Vainshtein1978, Zeldovich1987, Dolginov1978}. In particular, when constructing a dynamo model, it is necessary to account for the internal structure of the astrophysical object, which is typically not known with sufficient accuracy. These features include, for example, the pattern of internal differential motions, the conductivity of the material, the distribution of hydrodynamic quantities, and others. Moreover, for the dynamo to operate, an initial seed field is required, which must be generated by some other mechanism.

Thus, there is currently no unified dynamo theory. Instead, there are several not entirely consistent dynamo models, such as the Rikitake dynamo \cite{Rikitake1958}, the Faraday disk (unipolar induction), and others. The dynamo theory can explain the main characteristics of the magnetic fields of the Earth, the Sun, and other planets in the solar system, but only at a qualitative level. This is primarily due to the fact that convective motions of a rotating conducting fluid in the interiors of these objects represent an extremely complex phenomenon and are specific to each object \cite{Busse1976, Stevenson1983, Muzitani1992, Sano1993}. Dynamo models use many free parameters, the values of which are often poorly known or entirely unknown. Therefore, such models require additional calibration or tuning based on experimental data. As a result, dynamo models that well explain the features of the Earth's magnetic field show significant discrepancies with observational data for other planets. For example, the actual magnetic field of Mercury turned out to be 30 times weaker than theoretically predicted \cite{Ness1975}. In the case of Venus, theoretical predictions overestimated the magnetic field by three orders of magnitude compared to measurements \cite{Christensen2006}.

Another important point, considering the ideas presented in this work, is the necessity of deviation from axial symmetry \cite{Cowling1933} for the dynamo generation of a planet's magnetic field. This can be ensured by Coriolis forces in rotating systems, leading to the inclination of the planet's magnetic dipole axis relative to the rotation axis by a certain angle \cite{Stevenson1983}. However, this angle can change over time. For example, the Earth's magnetic field exhibits quasi-periodic fluctuations with characteristic times of $10^3$--$10^4$ years. If the magnetic field is averaged over these fluctuations, the average dipole will be oriented along the rotation axis.

Thus, a quantitative description of the magnetic fields of real objects within the framework of kinematic dynamo theory faces fundamental difficulties. Although there are currently no clear examples of celestial bodies whose magnetic fields cannot be explained by dynamo theory, the relevance of studying mechanisms of ''other nature'' is evident.

Several authors have proposed alternative hypotheses for the origin of magnetic fields, based on the separation of electric charges of electrons and atomic nuclei. For example, the idea of field generation due to the Hall current was developed by Vestine \cite{Vestine1954}. Other ideas include the use of the Nernst effect proposed by Gunn \cite{Gunn1929} (1936, unpublished), the excitation of electric currents under pressure effects by Inglis \cite{Inglis1955}, and others. It should be noted that these models also relied on not always clear mechanisms of charge formation and separation, the diurnal rotation of which would provide the initial field, subsequently amplified by the galvanomagnetic effect (Hall effect). Currently, ideas about certain physicochemical processes leading to charge separation are represented by V.V. Kuznetsov \cite{Kuznetsov2008} (thermodiffusion separation) and the works of V.I. Grigoriev et al. \cite{Grogoriev2003} (baroelectric effect).

The goal of this work is to continue the study, within the framework of a series of publications \cite{Vladimirov2000, Vladimirov2018, Babenko2020, Babenko2021}, of the possibility of a geometric approach \cite{geomphys} to explaining the origin of magnetic fields of astrophysical objects. For this purpose, we consider the hypothesis proposed in the early 20th century by Sutherland \cite{Sutherland1900a, Sutherland1900b, Sutherland1903, Sutherland1904, Sutherland1908} and Einstein \cite{Schwinger1986, Einstein1965}. Sutherland (1900--1908) put forward an unusual hypothesis about the origin of the Earth's magnetic field. He suggested that the observed magnetic field of the Earth is produced by the contributions of two oppositely directed magnetic fields generated by the rotations of: 1) the Earth's volumetric positive charge and 2) surface negative charges. It was assumed that the electric fields created by them are compensated, but this did not mean that their magnetic fields are also compensated. The resulting field was also thought to be responsible for the difference between the Earth's rotation axis and the magnetic axis, as well as for pole movement. The tilt of the magnetic axis was explained by the asymmetric distribution of conducting material, which also accounted for secular variations in the Earth's magnetic field.

In the pioneering works of Yu.S. Vladimirov \cite{Vladimirov2000, geomphys}, an explanation of the magnetic field of astrophysical objects was proposed, based on the hypothesis of charge separation (the author was unaware of the earlier hypothesis by Sutherland and Einstein) within the framework of multidimensional geometric models of physical interactions, such as Kaluza--Klein theories. In this case, Yu.S. Vladimirov derived a formula for the magnetic dipole moment of astrophysical objects based on the consequences of the Kerr--Newman metric and multidimensional Kaluza--Klein-type theories.

In this paper, which is an extended version of our recent work \cite{Babenko2025}, we briefly consider the theoretical part of the charge separation explanation within the 6D Kaluza--Klein theory. In Section 2, we derive the formula for the magnetic moment of a rotating gravitating body based on the consequences of the Kerr--Newman metric. In Section 3, we apply this formula to real astrophysical objects, specifically hot exoplanet giants belonging to the class of hot Jupiters \cite{hotjup-book}. The Conclusion summarizes the main findings of the work.

\section{Justification of Sutherland--Einstein hypothesis within 6D Kaluza--Klein theory and consequences of Kerr--Newman metric}

Many authors have noted the need for a proper theoretical justification of the Sutherland--Einstein hypothesis, i.e., explaining the difference in the magnitudes of the electric charges of the proton and electron. This can be done within the framework of 5D and 6D extensions of general relativity. Let us briefly consider the justification of the Sutherland--Einstein idea in a simplified version \cite{geomphys} of the 6D Kaluza--Klein theory \cite{Kaluza1921, Klein1926}.

In the literature, the term ''Kaluza--Klein theory'' is often incorrectly used to refer to 5D geometric models. In fact, there are two variants of 5D theories: first, the 5D Kaluza theory \cite{Kaluza1921}, aimed at geometrizing electromagnetism along with gravity, and second, the 5D Klein theory \cite{Klein1926}, intended for geometrizing particle masses. In Kaluza's theory, the wave functions of microparticles depend on the additional fifth coordinate $x^5$ as follows:
\beq\label{eq-2.1} 
 \Phi = \varphi(x^{\mu}) \exp \left( \frac{iec}{2\sqrt{G}\hbar} x^5 \right),
\eeq
where $\varphi(x^{\mu})$ describes the part of the wave function that depends only on the four classical space-time coordinates. In Klein's 5D theory, the wave functions of microparticles depend on another additional coordinate $x^4$ in a similar way, but with the particle mass replacing the charge:
\beq\label{eq-2.2}
 \Phi = \varphi(x^{\mu}) \exp \left( \frac{imc}{\hbar} x^4 \right).
\eeq

The unification of these two theories is achieved by increasing the dimensionality to six, with two additional coordinates $x^4$ and $x^5$. As a result, the dependencies on the additional coordinates \eqref{eq-2.1} and \eqref{eq-2.2} are combined. It is this theory that should be called the Kaluza--Klein theory. For the correct introduction of physically significant expressions, each of the two theories separately is constructed based on the monad method of 1+4-splitting, while the Kaluza--Klein theory is constructed based on the dyad method of 1+1+4-splitting. In such a theory, particle interactions are described using a dyadic operator of 4D differentiation, invariant under transformations of the two additional coordinates and covariant with respect to 4D transformations. This operator has the form:
\beq\label{eq-2.3} 
 \partial^{\dag\dag}_{\mu} =
 \frac{\partial}{\partial x^{\mu}} + 
 \left( \xi^4 \xi_{\mu} + \lambda^4 \lambda_{\mu} \right) 
 \frac{\partial}{\partial x^4} + 
 \lambda^5\lambda_{\mu} 
 \frac{\partial}{\partial x^5},
\eeq
where the vector $\lambda_A$ is directed along the Kaluza direction, and the vector $\xi_A$ along the Klein direction. The index $A$ runs from 0 to 5.

In the unified theory, the combination $\lambda^5\lambda_{\mu}$ is identified, as in the 5D Kaluza theory, with the vector potential of the electromagnetic field $A_{\mu}$. However, the expression above \eqref{eq-2.3} contains another combination $\xi^4\xi_{\mu} + \lambda^4\lambda_{\mu}$, which depends on the components of the multidimensional metric tensor $G_{4\mu}$ and actually describes some additional Abelian gauge field. In the simplified version of the theory under consideration, it is also proposed to identify it with the vector potential of the electromagnetic field \cite{geomphys}. This leads to the appearance of extremely small corrections in electromagnetic interaction due to the fact that the charge contribution arising during differentiation with respect to $x^4$ (actually proportional to the particle mass) is many orders of magnitude smaller than the nominal charge value.

It is easy to show that within the framework of the considered 6D geometric model, the mass $m$ induces an additional (''mass'') electric charge:
\beq\label{eq-dq}
\Delta q = 2\sqrt{G}m.
\eeq 
For example, for an electron with mass $m_\text{e} \approx 9.1 \cdot 10^{-28}$ g, the ratio of the additional charge to the main one is:
\beq\label{eq-dq2}
 \frac{\Delta e}{e} \approx 10^{-21}. 
\eeq
This conclusion corresponds to the Sutherland--Einstein hypothesis about the presence of a small charge asymmetry of elementary particles.

Obviously, such a correction to the electromagnetic interaction of particles lies beyond the accuracy of laboratory experiments. However, for large electrically quasi-neutral masses, where the electric charges of particles of opposite signs are on average compensated, the ''mass contribution'' \cite{Vladimirov2000, geomphys} to electromagnetic interaction turns out to be significant.

Let us consider massive astrophysical objects such as planets and stars. It is believed that on average such objects are electrically neutral, but according to the above, their charge should not be zero. Therefore, the space-time around spherical rotating charged objects should be described by a Kerr--Newman-type metric \cite{Newman1965}, where the additional constant (electric charge) is expressed through the mass value according to equation \eqref{eq-dq}. The additions to general relativistic effects in this case are extremely small, but both electric and magnetic fields should arise around such objects.

The Kerr--Newman geometry and the electromagnetic field associated with this one arise as a result of jointly solving the coupled Einstein--Maxwell equations under the conditions imposed by the quantities $M$ (mass), $q$ (charge), $L$ (intrinsic angular momentum), and the existence of a horizon. In the Kerr--Newman metric, written in Boyer--Lindquist coordinates ($t$, $r$, $\theta$, $\varphi$) \cite{Boyer1967}, which are a generalization of Schwarzschild coordinates, the square of the four-dimensional interval between two infinitely close events has the form:
\bearr\label{eq-ds}
 ds^2 = 
 \frac{\Delta}{p^2}
 \left( c dt - a \sin^2\theta d\varphi \right)^2 - 
\nnn
 - \frac{\sin^2\theta}{p^2}
 \left[ a c dt - (r^2+a^2) d\varphi \right]^2 - 
\nnn
 - \frac{p^2}{\Delta} dr^2 - 
 p^2 d\theta^2,
\ear
where the following notations are used:
\beq\label{eq-2.4a}
 \Delta = r^2 - r_\text{g} r + r^2_q + a^2,
\eeq
\beq\label{eq-2.4b}
 p^2 = r^2 +a^2 \cos^2\theta,
\eeq
\beq\label{eq-2.5}
 a = \frac{L}{Mc}, \quad 
 r_\text{g} = \frac{2GM}{c^2}, \quad
 r_q^2 = \frac{q^2G}{c^4}.
\eeq
The quantity $a$ represents the angular momentum parameter, $r_\text{g}$ is the gravitational radius, and $r_q$ is the characteristic radius due to the electric charge. The components of the vector potential $A_\mu$ and the electromagnetic field tensor $F_{\mu\nu}$ are determined by the expressions:
\bearr\label{eq-A}
 A_t = \frac{qr}{p^2}, \quad 
 A_r = A_\theta = 0, \nnn
 A_\varphi = -\frac{qar}{p^2} \sin^2\theta,
\ear
\beq\label{eq-F01}
 F_{tr} = \frac{q}{p^4} \left( r^2 - a^2 \sin^2\theta \right), 
\eeq
\beq\label{eq-F02}
 F_{t\theta} = \frac{q}{p^4} \left( r^2 - a^2 \sin^2\theta \right), 
\eeq
\beq\label{eq-F13}
 F_{r\varphi} = \frac{qa}{p^4} \sin^2\theta \left( r^2 - a^2 \sin^2\theta \right), 
\eeq
\beq\label{eq-F23}
 F_{\theta\varphi} = -\frac{2qar}{p^4} \sin\theta \cos\theta \left( a^2 + r^2 \right).
\eeq
The remaining components of the electromagnetic field tensor are either zero or obtained from these by changing the sign.

At large distances, when $r \gg \max(r_\text{g}, r_{q})$, space-time becomes almost flat. In a spherical coordinate system in an orthonormal basis, the components of the electric and magnetic fields take the form:
\bearr\label{eq-EB}
 E_{\hat{r}} = \frac{q}{r^2}, 
\nnn
 B_{\hat{r}} = \frac{2qa}{r^3} \cos\theta, \quad 
 B_{\hat{\theta}} = \frac{qa}{r^3} \sin\theta.
\ear
From this, it can be seen that the Kerr--Newman solution corresponds to an object of mass $M$, charge $q$, and angular momentum $L = aMc$, with the magnetic dipole moment being $\mu = qa$.

The value of the magnetic moment, as well as other properties of the astrophysical object described by the Kerr--Newman metric, are uniquely determined by the mass, charge, and angular momentum of the object. This corresponds to the ''no-hair theorem'', which states that all stationary black hole solutions of the Einstein--Maxwell equations of gravity and electromagnetism in general relativity can be fully characterized by only three independent externally observable classical parameters: mass, angular momentum, and electric charge. Other characteristics, such as geometry and magnetic moment, are uniquely determined by these three parameters.

In this paper, we consider the property of the electromagnetic field at large distances from the source when $r \gg \max(r_\text{g}, r_{q})$. From the point of view of the 6D Kaluza--Klein theory, the charge $q$ of a gravitating body is induced by its mass $M$ and is determined by the expression \eqref{eq-dq}. Then the dipole magnetic moment is:
\beq\label{eq-mu1}
 \mu = \frac{2\sqrt{G}L}{c}.
\eeq
Let us consider as a source a homogeneous sphere of radius $R$ rotating with angular velocity $\Omega$. In this case, the moment of inertia relative to the axis passing through its center is $I = \frac{2}{5} MR^2$, and the corresponding angular momentum is $L = \frac{2}{5} MR^2\Omega$. Substituting these expressions into the formula for $\mu$, we obtain:
\beq\label{eq-mu}
 \mu = \frac{4\sqrt{G}}{5c} M R^2 \Omega.
\eeq
This relationship, obtained within the framework of the geometric approach based on the 6D Kaluza--Klein theory, describes a certain fundamental magnetic field of any massive rotating body.

It should be emphasized that the formula \eqref{eq-mu} is valid only in some ideal situation. In real astrophysical objects such as planets and stars, the redistribution of charges and currents will inevitably occur. As a result, the fundamental field \eqref{eq-mu} will be largely compensated by absorbed (surface and volume) charges of the opposite sign. In addition, internal currents, due, for example, to differential rotation, will lead to the generation of an additional magnetic field. However, in the general case, this does not mean that the fundamental magnetic field will be completely compensated.

It should be expected that the magnetic field of such objects will consist of two parts: 1) the primary magnetic field, due to the additional electric charge \eqref{eq-dq}, and 2) the secondary magnetic field created by surface charges and internal currents. As a result, the resulting magnetic field will depend on the physical and chemical conditions affecting the distribution of absorbed charges, as well as the power of the conducting layer. The ratio between these fields in different astrophysical objects can vary greatly. For example, for the Earth, the fundamental field exceeds the observed one by almost 14 times, and for Jupiter, by 78 times. Hence, it makes sense to use the formula only for objects with approximately the same mechanism of generating their own field.

\section{Magnetic fields of hot Jupiters}

As an application of the described theory, let us consider exoplanets belonging to the type of hot Jupiters \cite{hotjup-book}. These planets are gas giants with mass $M > 0.25 M_{\rm J}$, where $M_{\rm J}$ is the mass of Jupiter, and their orbits are located close to the host star (semi-major axis $A < 0.1$ AU). Hot Jupiters are convenient for research because they are relatively easy to observe during transit--passing across the disk of the host star. This circumstance is due to two main factors. First, a typical hot Jupiter has a sufficiently large radius, resulting in a noticeable weakening of the star's brightness during transit. Second, the orbital period of these planets is short, and therefore transits occur quite frequently. The first hot Jupiter was discovered in 1995 \cite{Mayor1995}. As of the writing of this article (June 2025), 709 hot Jupiters are known\footnote{\url{http://www.exoplanet.eu}}.

Due to strong tidal effects, the proper rotation of a hot Jupiter becomes synchronized with its orbital rotation. This means that the planet's rotation period around its own axis becomes equal to the orbital rotation period. In addition, the action of tidal forces leads to a significant weakening of differential rotation in the planet's interior. In this state, the dynamo process of magnetic field generation becomes inefficient. The upper layers of the hot Jupiter's atmosphere can also generate a magnetic field \cite{Rogers2017}, as they consist of gas ionized by the hard ultraviolet radiation of the host star. However, such a field has a more complex configuration compared to the dipole field and decreases rapidly with distance. Some contribution to the planetary magnetic field can also be made by induced currents in the extended envelope of the hot Jupiter. These currents arise during the interaction of the planet with the stellar wind. A similar effect is observed, for example, in the magnetospheres of Venus and Mars. In the case of hot Jupiters, induced currents can also arise due to direct interaction with the magnetic field of the host star if the planet's orbit falls within the corona region.

Various methods for determining the magnetic field of these planets can be found in the monograph \cite{Lammer2015}. It should be emphasized that all these methods are model-dependent to some extent and do not allow an unambiguous determination of the magnetic field value. Therefore, the question of the magnitude and configuration of the magnetic field of hot Jupiters remains open. Researchers' opinions are conditionally divided into two groups. Some adhere to the conservative view that the magnetic field of hot Jupiters should be relatively weak. For example, observational estimates for the planet HD 209458 b \cite{Kislyakova2014} give a characteristic magnetic moment value $\mu/\mu_{\rm J} = 0.1$, where $\mu_{\rm J} = 1.53 \cdot 10^{30}$ G $\cdot$ cm$^3$ is the magnetic moment of Jupiter. In a number of other works (see e.g. \cite{Christensen2009, Vidotto2010, Ben-Jaffel2022}), the authors find much stronger magnetic fields. For example, for the planets HD 189733 b, HD 179949 b, $\tau$ Boo b, and $\upsilon$ And b, magnetic moments $\mu/\mu_{\rm J}$ of 3.0, 14.7, 16.4, and 15.9, respectively, were obtained \cite{Cauley2019}. The methodology for determining the magnetic field used by these authors raises a number of questions. Moreover, no mechanism is proposed that would lead to the generation of such a strong field. Therefore, in this work, we will adhere to the conservative assumption of a relatively weak magnetic field of hot Jupiters.

Since the efficiency of the dynamo in these planets sharply decreases due to strong tidal effects, it is reasonable to assume that the magnetic fields of these astrophysical objects are due to the same universal mechanism. Let us assume that the magnetic moment of a hot Jupiter is $\mu = x \mu_{\rm GR}$, where $\mu_{\rm GR}$ is determined by the expression \eqref{eq-mu}, and $x$ is some coefficient, the same for all planets of the considered type. This simple formula can be used to estimate the magnetic field of any hot Jupiter. However, to obtain an unambiguous value, it is necessary to calibrate this relationship by specifying a specific value of the coefficient $x$. For this, it is sufficient to take the value of the magnetic moment obtained by an independent method for some hot Jupiter. As such an independent estimate, we will use the results obtained in the work based on observational data for the planet HD 209458 b \cite{Kislyakova2014}. The field value of this hot Jupiter was $\mu = 0.1 \mu_{\rm J}$, which gives for our calibration the coefficient $x = 0.0084$.

\begin{figure*}
\centering
\includegraphics[width=\textwidth]{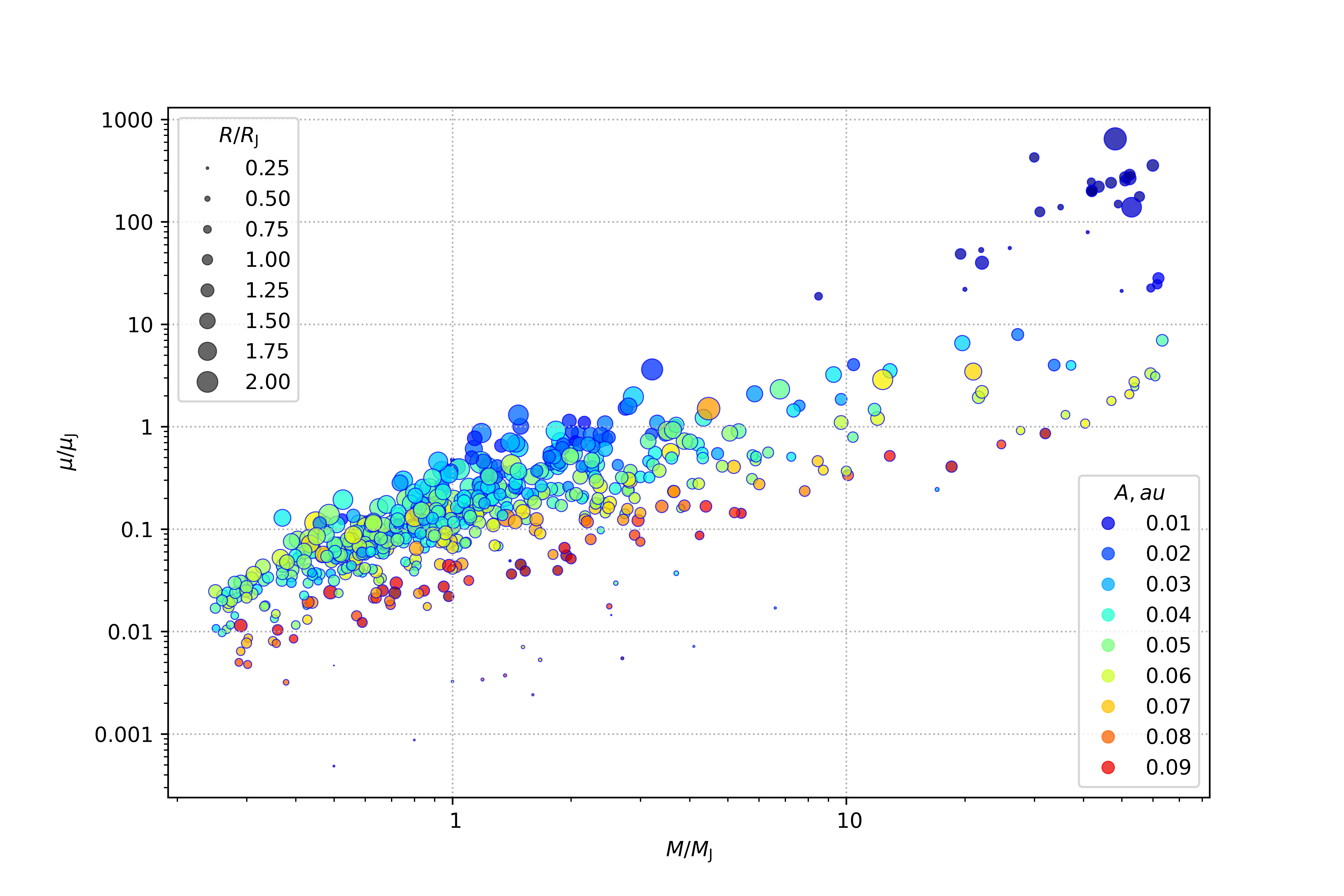}
\caption{\small 
Distribution of magnetic moments of hot Jupiters by their masses. The size of the circles corresponds to the radius of the planet, and the color --- the semi-major axis of the orbit.}
\label{fg-hjmf1}
\end{figure*}  

Fig. \ref{fg-hjmf1} shows the distribution of the magnetic moments $\mu$ of hot Jupiters calculated by mass $M$. The size of the circles shows the radius of the planet $R$, and the color scale corresponds to the semi-major axis. Our sample contains 656 objects, since it includes only those planets for which all the necessary parameters are known (mass $M$, radius $R$, orbital period $P_\text{orb}$). Hot Jupiters were considered to be in a state of synchronous rotation, when the period of their proper rotation is exactly equal to the orbital period. Therefore, the angular velocity of the planet's rotation around its own axis was set equal to $\Omega = 2\pi/P_\text{orb}$.

As can be seen from the figure, the magnetic moments of hot Jupiters lie in a fairly wide range of values, occupying 6 orders of magnitude, $10^{-3} \le \mu/\mu_\text{J} \le 10^3$. Planets with anomalously low fields ($\mu < 0.01 \mu_\text{J}$) are compact hot Jupiters of small radius (less than half the radius of Jupiter). There are also hot Jupiters with sufficiently strong magnetic fields ($\mu > 10 \mu_\text{J}$). Such planets are characterized by short orbital periods and large masses. It should also be noted that for such planets our estimates of the magnetic field may not be entirely correct. The fact is that such ultra-hot Jupiters actually rotate in the corona of the parent star. Therefore, the magnetic fields in them can be significantly induced by the field of the star.

\begin{figure*}
\centering
\includegraphics[width=\textwidth]{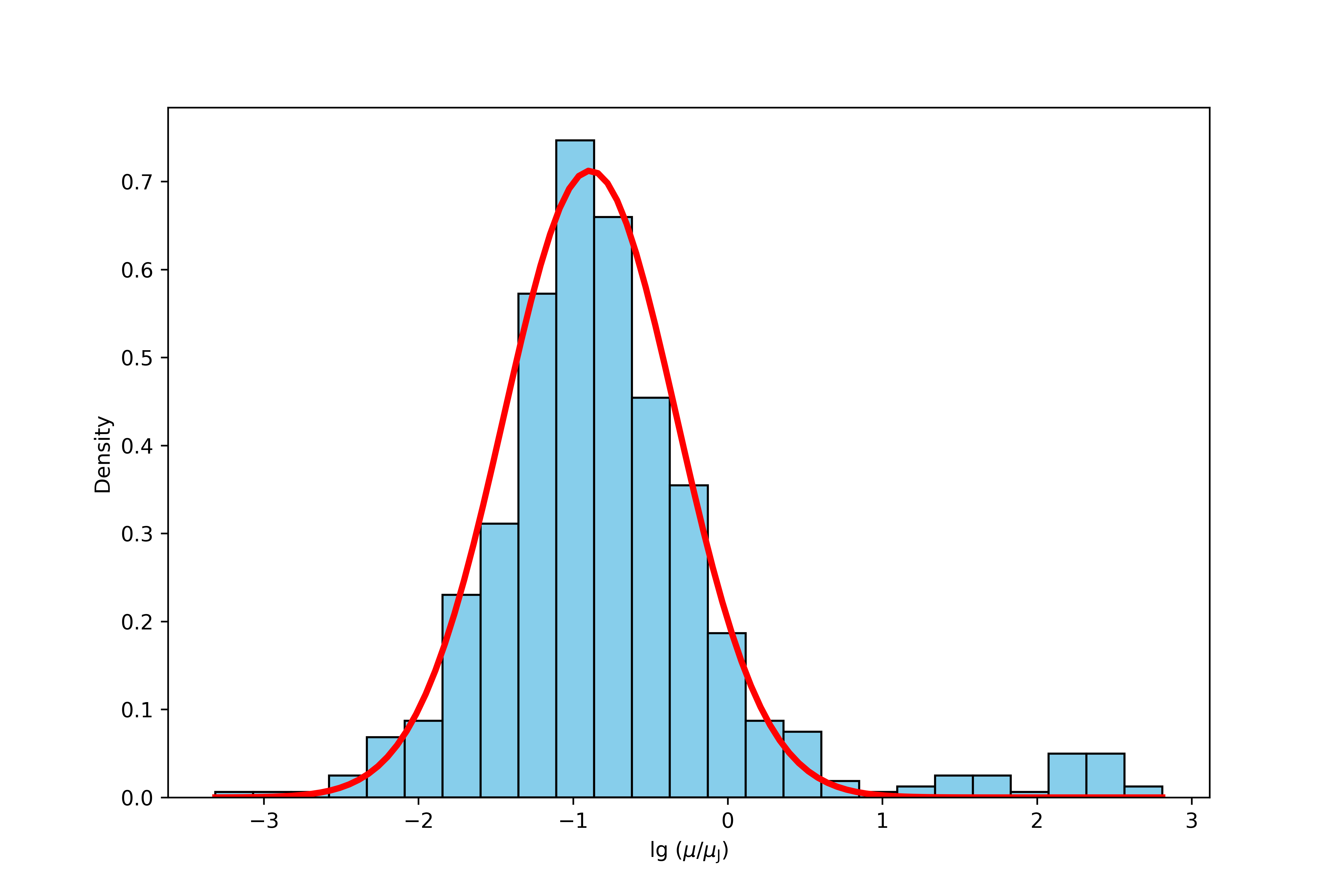}
\caption{\small 
Histogram of the distribution of hot Jupiters by magnetic moments. The solid curve shows a lognormal distribution with mean $a = -0.89$ and standard deviation $\sigma = 0.56$.}
\label{fg-hjmf2}
\end{figure*}  
   
The histogram of the distribution of hot Jupiters by magnetic moments is shown in Fig. \ref{fg-hjmf2}. Most of the planets from our sample have magnetic moments lying in the range of values $0.05 \le \mu/\mu_\text{J} \le 0.5$. We approximated this histogram with a lognormal distribution function (normal distribution from the value $\lg\mu$). The resulting function is shown in the figure by the solid curve. We obtained the following values of the mathematical expectation and standard deviation:
\beq 
 a = -0.89 \pm 0.02, \quad 
 \sigma = 0.56 \pm 0.02.
\eeq
The found mathematical expectation $a$ corresponds to the average value of the magnetic moment $\bar{\mu} = 0.13 \mu_\text{J}$.

\section{Conclusion}

The original idea for our research was the Sutherland--Einstein hypothesis about the cause of the emergence of magnetic fields in astrophysical objects. The authors of this hypothesis suggested that the magnetic field in such bodies arises due to a very small charge asymmetry of the proton and electron. Massive astrophysical objects such as planets or stars are on average electrically neutral, since their additional electric charge, due to the difference in the charges of the electron and proton, is compensated by absorbed charges of the opposite sign. However, the magnetic field arising from rotation will generally not be compensated. As a result, an effective magnetic field appears around such objects, representing a superposition of two oppositely directed parts: 1) the primary magnetic field of the additional electric charge due to mass and 2) the secondary magnetic field created by absorbed charges and currents. The resulting magnetic field depends significantly on the distribution of absorbed charges and internal currents.

In our work, the theoretical justification of the Sutherland--Einstein hypothesis was built within the framework of a geometric approach based on the synthesis of the 5D theories of Kaluza and Klein. This theory made it possible to obtain a formula for the subsequent empirical calculation of the magnetic moment of a certain type of astrophysical objects. In the simplified 6D version of this theory, it is possible to show that the mass of a gravitating object induces an additional electric charge. This conclusion fully corresponds to the Sutherland--Einstein hypothesis about the presence of a small charge asymmetry of elementary particles. At the same time, the magnitude of the additional charge for electrons and protons turns out to be extremely small and lies outside the resolution limits of modern equipment in laboratory conditions.

From the point of view of general relativity, the geometry of 4D space-time around idealized rotating bodies should be described by the Kerr--Newman metric. However, the additional constant (electric charge) will now be expressed directly through the mass value. As a result, both electric and magnetic fields will arise around such massive rotating objects. Using the formula for the dipole magnetic moment of the Kerr--Newman source, one can obtain an expression for the primary magnetic field of such idealized objects.

For real planets and stars, the ratio of the primary magnetic field to the observed one varies greatly. However, this coefficient can take approximately the same values for a certain group of astrophysical objects, which are characterized by the same mechanism for generating their own compensating field. As an application of our theory, we considered such astrophysical objects as hot Jupiters. These exoplanets, which are gas giants, are located close to their host stars. As a result of strong tidal influence, their own rotation becomes synchronized with the orbital one. This means that differential rotation in the interior is very insignificant and, therefore, the dynamo mechanism of magnetic field generation becomes inefficient.

Assuming that for all exoplanets of this type, the compensation of the primary field occurs in the same way, we find:
\beq\label{eq-mu2} 
 \frac{\mu}{\mu_\text{J}} = 0.27\,  
 \frac{M}{M_\text{J}}
 \left( \frac{R}{R_\text{J}} \right)^2 
 \frac{\text{day}}{P_\text{orb}}.
\eeq
This formula allows us to calculate the magnetic moment of any hot Jupiter from the known observed values of mass $M$, radius $R$, and orbital period $P_\text{orb}$ (or semi-major axis). Our analysis leads to the conclusion that the magnetic moments of hot Jupiters are scattered over a fairly wide range of values, but for most planets $0.05 \le \mu/\mu_\text{J} \le 0.5$. At the same time, the average value of the magnetic moment is $\bar{\mu} = 0.13 \mu_\text{J}$, which agrees well with the conservative estimates of other authors. It is also interesting to note that for Jupiter equation \eqref{eq-mu2} gives $\mu = 0.66 \mu_{\rm J}$, which is quite close to the actual value. This result quantitatively substantiates the Sutherland--Einstein hypothesis. 

The rapid development of exoplanet research, in particular, opens up new possibilities for forecasting space weather. In the work \cite{fizmat2023}, a method for obtaining quantitative data on the parameters of stellar wind and coronal mass ejections through transit observations of the reaction of hot exoplanet atmospheres to the activity of their host stars is proposed. This will complement measurements in the Solar System, significantly enriching the statistical and evolutionary aspects of extreme manifestations of space weather. Since the planet's magnetic field plays an important role in these processes, the results of this work can make a useful contribution to these studies.


\Funding{This work was supported by program 10 ''Experimental laboratory astrophysics and geophysics'' of the National Center for Physics and Mathematics.}


\end{document}